\documentclass[twocolumn,prb,amsmath,amssymb,floatfix]{revtex4}

\usepackage{bbold}
\usepackage{bm}
\usepackage{color}
\usepackage{dcolumn}
\usepackage{feynmf} 
\usepackage{graphics}
\usepackage{graphicx}
\usepackage{subfigure}
\usepackage{slashed}
\usepackage{tensor}
\usepackage{placeins}
\usepackage{natbib}

\def\B{\langle }
\def\K{\rangle }

\def\Tr{\mbox{Tr}}

\def\al{\alpha}

\def\veps{\varepsilon}

\def\be{\begin{equation}}
\def\ee{\end{equation}}
\def\bea{\begin{eqnarray}}
\def\eea{\end{eqnarray}}
\def\bse{\begin{subequations}}
\def\ese{\end{subequations}}
\def\bc{\begin{center}}
\def\ec{\end{center}}

\def\ra{\rightarrow}

\def\nonum{\nonumber}

\def\I{{\rm i}}

\def\D{{\rm d}}
\def\Ord{{\rm O}}

\def\Sp{{\slashed p}}

\newcommand{\comment}[1]{}

\catcode`,\active

\catcode`\,12

\begin{document}

\title{ Interaction corrections to the minimal conductivity of graphene \\ via dimensional regularization}


\author{S.~Teber$^{1,2}$ and A.~V.~Kotikov$^3$}
\affiliation{
$^1$Sorbonne Universit\'es, UPMC Univ Paris 06, UMR 7589, LPTHE, F-75005, Paris, France.\\
$^2$CNRS, UMR 7589, LPTHE, F-75005, Paris, France.\\
$^3$Bogoliubov Laboratory of Theoretical Physics, Joint Institute for Nuclear Research, 141980 Dubna, Russia.}
\date{\today}

\begin{abstract}
We compute the two-loop interaction correction to the minimal conductivity of disorder-free intrinsic graphene with the help of
dimensional regularization. The calculation is done in two different ways: via density-density and via current-current correlation functions.
Upon properly renormalizing the perturbation theory, in both cases, we find that: $\sigma = \sigma_0\,( 1 + \al\,(19-6\pi)/12) \approx \sigma_0 \,(1 + 0.01\, \al)$, 
where $\al = e^2 / (4 \pi \hbar v)$ is the renormalized fine structure constant and $\sigma_0 = e^2 / (4 \hbar)$. Our results are
consistent with experimental uncertainties and resolve a theoretical dispute.
\end{abstract}

\maketitle


{\it Introduction} - Graphene is a one-atom thick layer of graphite, see, {\it e.g.}, Ref.~[\onlinecite{KotovUPGC12}] for a review,
where the quasiparticle spectrum is Dirac-like and massless at low-energies.~\cite{Wallace47,Semenoff84} 
Remarkably, despite the fact that disorder-free intrinsic graphene has a vanishing density of states at the Fermi points, the chiral nature of the charge carriers yields
a minimal ac conductivity: $\sigma_0 = e^2 / (4 \hbar)$, which is universal.
This result, which was predicted to hold for free Dirac fermions~\cite{LudwigFSG1994}, agrees to within $1$-$2\%$ with optical experiments.~\cite{C-experiments}
This is rather surprising because the long-range Coulomb interaction among charge carriers is unscreened and supposed to be strong; 
the fine structure constant of graphene, $\al = e^2 / (4 \pi \hbar v) \approx 2.2$, due to the fact that the Fermi velocity is smaller than the velocity of light, $v \approx c / 300$.

There has been extensive theoretical attempts to understand the effect of electron-electron interactions on the
homogeneous optical conductivity of graphene, see, {\it e.g.}, 
Refs.~[\onlinecite{Mishchenko2008,HerbutJV2008,JurcicVH2010,SheehyS2009,AbedinpourVPPTM2011,SodemannF2012,Rosenstein2013,GazzolaCCNS2013}].
The latter can be defined via a density-density correlation function:
\be
\sigma(q_0) = - \lim_{\vec{q} \ra 0} \, \frac{\I q_0}{|\vec{q}\,|^2}\,\Pi^{00}(q_0,\vec{q}\,)\, ,
\label{sigma-dd}
\ee
where $\Pi^{0 0} (q) = \B T \rho(q) \rho(-q) \K$, and $\rho$ is the charge density. Equivalently, from current conservation, it can also be defined 
via a current-current correlation function (Kubo formula):
\be
\tilde{\sigma}(q_0) = \frac{1}{\I q_0}\,\frac{K^{11}(q_0,\vec{q}=0\,) + K^{22}(q_0,\vec{q}=0\,)}{2}\, ,
\label{sigma-cc}
\ee
where $K^{i j} (q) = \B T j^i(q) j^j(-q) \K$ and $\vec{j}$ is the charge current.

\begin{figure}[ht]
    \includegraphics{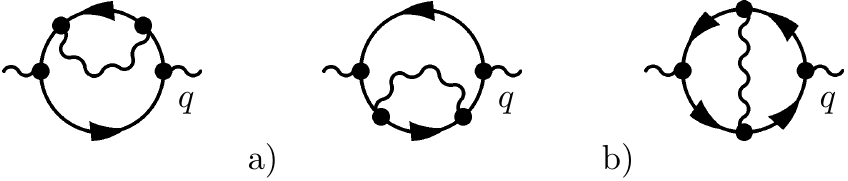}
    \caption{\label{fig:2loop-polarization}
     Two-loop vacuum polarization diagrams.}
\end{figure}

Despite the strength of the interactions, one may first focus on the lowest order interaction corrections to $\Pi^{00}(q)$ and $K^{ij}(q)$, see Fig.~\ref{fig:2loop-polarization}, from which:
\be
\sigma(q_0) = \sigma_0\, \bigg( 1 + \mathcal{C} \al + \Ord(\al^2) \bigg)\, ,
\ee
and equivalently for $\tilde{\sigma}$, and extract the numerical value of the first order interaction-correction coefficients, $\mathcal{C}$ and $\tilde{\mathcal{C}}$, respectively.
On physical grounds, one expects that $\mathcal{C} = \tilde{\mathcal{C}}$, independent on the method used.

\FloatBarrier

It turns out, however, that different theoretical results can be found in the literature so that the value of the first order coefficient is controversial. 
As quoted by Ref.~[\onlinecite{Rosenstein2013}], these results read:
\begin{subequations}
\label{c123}
\bea
\mathcal{C}^{(1)}  = \frac{25-6\pi}{12} \approx 0.512\, ,
\label{c1}\\
\mathcal{C}^{(2)}  = \frac{19-6\pi}{12} \approx 0.013\, ,
\label{c2}\\
\mathcal{C}^{(3)}  = \frac{11-3\pi}{6} \approx 0.263\, .
\label{c3}
\eea
\end{subequations}

Starting from the Kubo formula, Herbut et al.\ obtained:~\cite{HerbutJV2008}
$\tilde{\mathcal{C}}^{(\Lambda)}  = \mathcal{C}^{(1)}$.
%
%
They used a hard cut-off in order to regularize the UV-divergences arising from individual two-loop diagrams
and which ultimately cancel out in their sum.
On the other hand, starting from Eq.~(\ref{sigma-dd}), Mishchenko obtained:~\cite{Mishchenko2008}
\be
\mathcal{C}^{(\Lambda)} = \mathcal{C}_{a}^{(\Lambda)} + \mathcal{C}_{b}^{(\Lambda)} = \mathcal{C}^{(2)},
\quad \mathcal{C}_{a}^{(\Lambda)}  = \frac{1}{4}, \,\, \mathcal{C}_{b}^{(\Lambda)}  = \frac{8-3\pi}{6}\, ,
\label{c}
\ee
where individual two-loop diagrams are finite in this case and the hard cut-off only regularizes the divergent self-energy subgraph
of the two diagrams in Fig.~\ref{fig:2loop-polarization}a (the latter contribute to the value $\mathcal{C}_{a}$ while 
$\mathcal{C}_{b}$ comes from the diagram in Fig.~\ref{fig:2loop-polarization}b). 
Still using a hard cut-off, a third result, $\mathcal{C}_{kin}^{(\Lambda)} = \mathcal{C}^{(3)}$, was even obtained with the help of a kinetic equation approach.~\cite{Mishchenko2008}
%
%
Mishchenko claimed that these discrepancies are due to the long-range nature of the Coulomb interaction. He advocated the use of a soft cut-off in order to properly regularize
the UV-divergent integrals finding a single result given by $\mathcal{C}^{(2)}$.~\cite{Mishchenko2008}

In Ref.~[\onlinecite{JurcicVH2010}] (henceforth referred to as JVH) the coefficient was recomputed with the help of dimensional regularization (DR).
Starting from Eq.~(\ref{sigma-dd}), JVH obtained:
\be
\mathcal{C}^{(D)} =  \mathcal{C}_{a}^{(D)} + \mathcal{C}_{b}^{(D)} = \mathcal{C}^{(3)}\, \quad \mathcal{C}_{a}^{(D)}  = \frac{1}{2}, \,\, \mathcal{C}_{b}^{(D)}  = \frac{8-3\pi}{6}\, ,
\label{c-D}
\ee
in disagreement with Eq.~(\ref{c}). They obtain the same result via the Kubo formula: $\tilde{\mathcal{C}}^{(D)} = \mathcal{C}^{(3)}$, in disagreement with both
Eq.~(\ref{c}) and their previous result~\cite{HerbutJV2008} $\tilde{\mathcal{C}}^{(\Lambda)}  = \mathcal{C}^{(1)}$. 
Recently, support in favor of JVH's result, $\mathcal{C}^{(3)}$, came from yet another approach based on a full tight-binding computation [\onlinecite{Rosenstein2013}].
However, the most commonly accepted result is, up to date, the value $\mathcal{C}^{(2)}$ since it has been recovered by a majority of groups, 
mainly using hard cut-off regularization and variants of it, see, {\it e.g.}, Refs.~[\onlinecite{SheehyS2009,AbedinpourVPPTM2011,SodemannF2012,GazzolaCCNS2013}]. 
Incidentally, this is also the only result, among those of Eqs.~(\ref{c123}), which is consistent with the experimental uncertainties.~\cite{C-experiments}

In the present Letter we reconsider the computation of the minimal conductivity with the help of dimensional regularization.
Upon properly renormalizing the theory, we find that:
\be
\mathcal{C}^{(\rm{DR})} = \tilde{\mathcal{C}}^{(\rm{DR})} = \mathcal{C}^{(D)} + \mathcal{C'}^{(D)} = \mathcal{C}^{(2)}\, \quad \mathcal{C'}^{(D)} = -\frac{1}{4}\, ,
\label{c-DR}
\ee
where $\mathcal{C'}^{(D)}$ originates from one-loop counterterms. 
This suggests that, as far as DR is concerned, the origin of the controversy does not lie in the regularization method or in the possible presence of an anomaly
but, more simply, in the renormalization procedure itself. 

{\it Master integrals -} 
Contrary to JVH, who introduced Feynman parameters, we shall compute the multi-loop dimensionally regularized integrals using algebraic methods, therefore
providing an independent check of the calculations. The implementation of these methods requires the knowledge of some basic integrals
such as the massless one-loop propagator-type integral with $n\leq 2$
[\onlinecite{Kazakov:1986mu}]
%
\begin{subequations}
\label{massless-p-integral}
\bea
&&\int [\D^{D} q]\,\frac{q^{\mu_1}\dotsc q^{\mu_n}
}{[q^2]^{\al}[(q-k)^2]^\beta}  =
\frac{(k^2)^{D/2-\al-\beta}}{(4\pi)^D}\, 
\nonumber \\
&&\times \left[k^{\mu_1} \dotsc k^{\mu_n}
\, G^{(n,0)}_0(\al,\beta)\, + \delta^2_n \frac{g^{\mu_1 \mu_2}}{D} 
\, G^{(1,1)}_1(\al,\beta)\right], ~~~
\label{massless} \\
&&G^{(n,m)}_i(\al,\beta) = \frac{a_n(\al)a_m(\beta)}{a_{n+m-i}
(\al+\beta-D/2-i)}, 
\\
&&a_n(\al) = \frac{
\Gamma(D/2-\al + n)
}{\Gamma(\al)} \, ,
\eea
\end{subequations}
where $[\D^D q] = \D^{D} q / (2\pi)^{D}$ and  $\delta^2_n$ is the Kronecker
symbol.
%
%
The simplified notation: $G(\al,\beta)=G^{(0,0)}_0(\al,\beta)$, will also be used.
As we shall see in the following, the computation of the ac conductivity involves semi-massive tadpole diagrams. 
In particular, the one-loop semi-massive tadpole diagram reads:
\begin{subequations}
\label{tadpole}
\bea
&&\int \frac{[\D^D k]}{[k^2]^{\al}[k^2+m^2]^\beta} 
=
\frac{(m^2)^{D/2-\al-\beta}}{(4\pi)^{D/2}}\,B(\beta,\al)\, ,
\\
&&B(\beta,\al) = \frac{\Gamma(D/2-\al)\,\Gamma(\al + \beta - D/2)}{\Gamma(D/2)\,\Gamma(\beta)}\, .
\eea
\end{subequations}
%
%
These formulas can be used to compute all required 2-loop semi-massive tadpole diagrams. As will be seen in the following,
the latter are of the form:
\bea
I_n(\al) &=&  \int  \frac{[\D^{D_e}k_1][\D^{D_e}k_2]\,(\vec{k}_1 \cdot \vec{k}_2\,)^n
[|\vec{k}_1 - \vec{k}_2\,|^{2}]^{-1/2}}
{[|\vec{k}_1\,|^2]^\al\,[|\vec{k}_1\,|^2+m^2]\,[|\vec{k}_2\,|^2]^{\al}\,[|\vec{k}_2\,|^2+m^2]
} \nonumber \\
&=&  \frac{(m^2)^{D_e+n-2\al-5/2}}{(4\pi)^{D_e}} \, \tilde{I}_n(\al)
\, .
\label{I(n)}
\eea
The diagrams with different $\al$ values are related to each other by the relation:
\be
\frac{1}{k^{2\al}\,[k^2+m^2]} =
\frac{1}{m^2} \left[ \frac{1}{k^{2\al}} -
\frac{1}{k^{2(\al-1)}\,[k^2+m^2]} \right] \, .
\ee
The diagrams $I_n(\al)$ with some particular $\al$ values can be calculated by 
Eqs.~(\ref{massless-p-integral})-(\ref{tadpole}) when one of the massive propagators
can be replaced by a massless one with the help of a Mellin-Barnes transformation [\onlinecite{Boos:1990rg}]:
\be
\frac{1}{k^2 + m^2} = \frac{1}{2\I \pi}\,\int_{-\I \infty}^{+\I \infty} \D s \,\Gamma(-s) \Gamma(1+s) \frac{(m^2)^s}{(k^2)^{1+s}}\, . 
\ee
The results have the following form for $\veps_\gamma \ra 0$:
\be
\tilde{I}_0(1/2)= -\tilde{I}_1(3/2)= \tilde{I}_2(5/2)= 
\,\pi^2\, ,
\label{masters}
\ee
We note that really only one of the diagrams, for example $I_0(1/2)$, is
independent. The two others can be expressed through $I_0(1/2)$ using
integration by parts identities.
However, this procedure is quite long
and we will show it in our future publication [\onlinecite{Kotikov:2014NEW}].
The diagrams with other  $\al$ values can be expressed as combinations
of the ones in Eq.~(\ref{masters}) and of simpler diagrams, which can be calculated  
directly with help of Eqs.~(\ref{massless-p-integral})-(\ref{tadpole}).
So, we have, for the diagrams contributing to $\Pi^{00}(q)$ at two loops (see 
Eq.(\ref{pi2b00-inter}) below) and $\veps_\gamma \ra 0$
\be
\tilde{I}_1(1/2)= \,\pi(4-\pi)\, ,~~ \tilde{I}_2(3/2)= \,\pi\left(\pi -
\frac{4}{3}\right) \, .
\label{master-dd}
\ee
The diagrams contributing to $\Pi(q^2)$ at two loops are shown below in
Eq.~(\ref{pi2bmumu-inter}). They are UV-singular and read:
\bea
&&\tilde{I}_1(-1/2)=  
\pi^2 - 2 G\left(\frac{1}{2},\frac{1}{2}\right)
\, B\left(1,\frac{3- D_e}{2}\right)
\, , \label{master-cc} \\
&&\tilde{I}_2(1/2)=  
\pi^2 - \frac{4\pi}{3} 
- (3-D_e)\,G\left(\frac{1}{2},\frac{1}{2}\right)
\,B\left(1,\frac{3- D_e}{2}\right)
\nonumber
\eea
with the accuracy $ O(\veps_\gamma)$.

{\it Feynman rules and renormalization -} The effective low-energy action of graphene, in Minkowski space, reads:
\bea
S = &&\sum_{n=1}^{N_F} \int \D t\, \D^2 x\, \left[ \bar{\psi}_n \left( \I \gamma^0 \partial_t - \I v_0 \vec{\gamma} \cdot \vec{\nabla}\,\right) \psi_n \right .
\nonum \\
&&\left . - e_0 \bar{\psi}_n \,\gamma^0 A_0\, \psi_n \right] + \frac{1}{2}\,\int \D t\, \D^3 x\,(\vec{\nabla} A_0)^2\, ,
\label{action}
\eea
where $v_0$ and $e_0$ are the bare Fermi velocity and charge, respectively,
$\psi_n$ is a four-component spinor field describing a fermion of specie $n$ (for graphene $N_F=2$) and
$A_0$ is the gauge field mediating the instantaneous Coulomb interaction.
The Dirac matrices, $\gamma^\mu = (\gamma^0,\vec{\gamma}\,)$ satisfy the usual algebra $\{\gamma^\mu,\gamma^\nu \} = 2g^{\mu \nu}$,
with metric tensor $g^{\mu \nu} = \text{diag}(+,-,-)$. 

From Eq.~(\ref{action}), the bare momentum space fermion propagator reads (we use the convention
$p^\mu=(p^0,v_0 \vec{p}\,)$):
\be
S_0(p) = \frac{i\Sp}{p^2}\, , \qquad \Sp = \gamma^\mu p_\mu = \gamma^0 p_0 - v_0 \vec{\gamma}\cdot \vec{p}\, ,
\label{fermion-prop0}
\ee
and we shall implicitly assume that Feynman's prescription, $p^2 \equiv p^2 + \I0^+$, holds so that we may Wick rotate all integrals.
The effective photon propagator reads:
\be
V_0(\vec{q}\,) = \frac{\I}{2 (|\vec{q}\,|^2)^{1/2}}\, ,
\label{gauge-field-prop0}
\ee
and the bare vertex is: $-\I e_0 \gamma^0$. 

In conventional DR,~\cite{Kilgore2011} these Feynman rules stay the same but momenta, Dirac matrices and metric tensor are extended to span a $D_e$-dimensional space
(keeping $\Tr \left[ \mathbb{1} \right] = 4 N_F$) with $D_e = 2 - 2\veps_\gamma$. 
All bare parameters and fields are then related to renormalized ones via renormalization group constants: 
$\psi_n = Z_{\psi}^{1/2}\psi_{nr}$, $A_0 = Z_A^{1/2} A_{0r}$ and $v_0 = Z_v v$.
We shall use DR in the $\overline{\rm{MS}}$ scheme, where the $Z$s are polynomial in $1/\veps_\gamma$ and the bare charge, $e_0$, is related to the renormalized one,
$e$, via:
\be
\frac{e_0^2}{(4\pi)^{D_e/2}} = \frac{e^2(\mu)}{4\pi}\,\mu^{2\veps_\gamma}\,Z_e^2(\mu)\,e^{\gamma_E \veps_\gamma} \, ,
\label{ren_coupling}
\ee
where $\mu$ is the renormalization scale. In graphene, charge does not flow and: $Z_e = 1$.

\begin{figure}
    \includegraphics{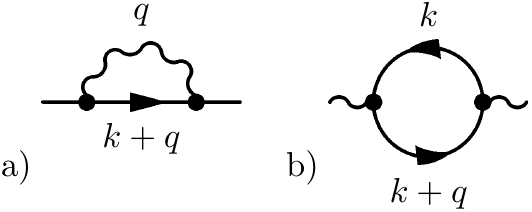}
  \caption{\label{fig:one-loop}
  One-loop a) fermion and b) photon self-energies.}
\end{figure}

We may now focus on the one-loop fermion self-energy, Fig.~\ref{fig:one-loop}a. The latter is defined as:
\be
-i\Sigma_1(k) = \int [\D^{d_e} q]\, (-\I e_0 \gamma^0) \,S_0(k+q)\, (-\I e_0 \gamma^0)\,V_0(q)\, ,
\nonum
\ee
where $d_e = 1 + D_e$ is the space-time dimension. Because of the instantaneous nature of the interaction this self-energy depends only on momentum; there is
therefore neither wave-function nor vertex renormalization: $Z_\psi = Z_\Gamma^{-1} = 1$.
The parametrization: 
\be
\Sigma_1(\vec{k}\,) =  v_0 \vec{\gamma} \cdot \vec{k}\,\Sigma_{k 1}(|\vec{k}\,|^2)\, , \quad
\Sigma_{k 1}(|\vec{k}\,|^2) = -\frac{\Tr[\vec{\gamma} \cdot \vec{k}\,\Sigma_1(\vec{k}\,)]}{4N_F v_0 |\vec{k}\,|^2}\, ,
\label{fsigma-param}
\ee
together with Eqs.~(\ref{massless-p-integral}), yields:
\be
\Sigma_{k1}(|\vec{k}\,|^2) =  \frac{e_0^2}{8\,v_0}\,\frac{(|\vec{k}\,|^2)^{D_e/2-1}}{(4\pi)^{D_e/2}}\,G(1/2,1/2)\,  .
\label{fsigma2}
\ee
Performing the $\veps_\gamma$-expansion in the $\overline{\rm{MS}}$ scheme, yields, with one-loop accuracy:
\be
\Sigma_{k1}(|\vec{k}\,|^{2}) = \frac{\al(\mu)}{8}\, \left( \frac{1}{\veps_\gamma} - L_k + 4 \log 2 + \Ord(\veps_\gamma) \right)\, ,
\label{fsigma2-exp}
\ee
where $\al$ is the renormalized coupling constant and $L_k = \log(|\vec{k}\,|^2/\mu^2)$. 
The UV-divergent self-energy leads to a renormalization of the Fermi velocity:~\cite{GonzalezGV93}
\be
Z_v = 1 - \frac{\al(\mu)}{8 \veps_\gamma} + \Ord(\al^2), \quad \al(\mu) = \frac{e^2(\mu)}{4\pi v(\mu)}\, .
\label{Zv}
\ee
The corresponding beta-function is negative: $\beta_v = \frac{\D \log v(e(\mu))}{\D \log \mu} = - \al/4$, 
implying that Fermi velocity grows in the infrared.~\cite{GonzalezGV93}

We may proceed in a similar way with the one-loop photon self-energy, Fig.~\ref{fig:one-loop}b, defined as:
\be
\I \Pi_1^{\mu \nu}(q) = - \int [\D^{d_e}k]\, \Tr \left[ (-\I e_0 \gamma^\mu)\,S_0(k+q)\,(-\I e_0 \gamma^\nu)\,S_0(k) \right] \, .
\label{pi1munu}
\ee
Focusing on $\Pi^{00}$, performing the trace, going to euclidean space ($q_0 = \I q_{E0}$),
integrating over frequencies and taking the $\vec{q} \ra 0$ limit, yields:
\be
\Pi_1^{00}(q_{E0},\vec{q} \ra 0) = \frac{N_F}{2v_0}\, e_0^2\,|\vec{q}\,|^2\,\frac{D_e-1}{D_e}\, 
\int \frac{[\D^{D_e}k]}{|\vec{k}\,|\, [|\vec{k}\,|^2 + m_0^2]}\, ,
\ee
which is of the form Eq.~(\ref{tadpole}) with $m_0 = q_{E0}/2 v_0$. This is immediately integrated to give:
\be
\Pi_1^{00}(q_{E0},\vec{q} \ra 0) = \frac{N_F}{2 v_0 m_0}\,|\vec{q}\,|^2\,\frac{e_0^2\,(m_0^2)^{-\veps_\gamma}}{(4\pi)^{D_e/2}}\,\frac{D_e-1}{D_e}\,B(1,1/2)\, .
\ee
Using Eqs.~(\ref{ren_coupling}) and (\ref{Zv}) to express the bare parameters in terms of renormalized ones 
and performing the $\veps_\gamma$-expansion yields, with two-loop accuracy:
\be
\Pi_1^{00}(q_{0},\vec{q} \ra 0) = -\frac{N_F e^2}{8}\,\frac{|\vec{q}\,|^2}{\I q_0}\,\left( 1 - \frac{\al}{4} \right)\, .
\label{pi100}
\ee
We note that Fermi velocity renormalization plays a crucial role in Eq.~(\ref{pi100}) as it brings the factor $(Z_v)^{2\veps_\gamma} = 1-\al/4$ to $\Ord(\veps_\gamma)$ accuracy.
Combining Eqs.~(\ref{sigma-dd}) and (\ref{pi100}), we arrive at: $\sigma_1(q_0) = \sigma_0\,(1 + \mathcal{C}'^{(D)} \al + \Ord(\al^2))$, with $\mathcal{C}'^{(D)} = -1/4$.

We may now proceed in a similar way with the help of the Kubo formula Eq.~(\ref{sigma-cc}). 
In order to better exploit the $O(2)$ space rotational symmetry of the system we shall derive an alternate formula based on Eq.~(\ref{pi1munu}).
Using the Ward identity, $S_0(k) (-\I e_0 \Sp) S_0(k+p) = e_0 [ S_0(k) - S_0(k+p) ]$, this function can be shown to be transverse: $q_\mu \Pi_1^{\mu \nu}(q) =0$, reflecting current conservation.
We note, from now on, that similar arguments apply to the 2-loop corrections, see Eqs.~(\ref{def:pi2loop}) below, so that: $q_\mu \Pi_2^{\mu \nu}(q) =0$, where $\Pi_2$ is the total two-loop contribution.
We may therefore attempt to parametrize $\Pi^{\mu \nu}$ as follows:
\be
\Pi^{\mu \nu}(q) = (g^{\mu \nu}q^2 - q^\mu q^\nu)\,\Pi(q^2), \quad
\Pi(q^2) = \frac{- \tensor{\Pi}{^\mu_\mu}(q)}{(d_e-1)(-q^2)}\, .
\label{def:pi2}
\ee
Then:
\be
\tilde{\sigma}(q_0) = \I q_0 \, K(q_0)\, ,
\label{sigma-cc2}
\ee
where $K(q_0) = v_0^2 \Pi(q_0^2,|\vec{q}\,|^2 \ra 0)$ and should be properly renormalized in the course of the computation.

Following the steps of the $\Pi^{00}$ computation, at one-loop, we have:
\be
K_1(q_{E0}) =  \frac{N_F}{2 v_0 m_0} \, \frac{e_0^2\,(m_0^2)^{-\veps_\gamma}}{(4\pi)^{D_e/2}}\,\frac{D_e-1}{D_e}\,B(1,-1/2)\, .
\ee
Expressing all bare parameters in terms of renormalized ones and performing the $\veps_\gamma$-expansion yields, with two-loop accuracy:
\bea
K_1(q_{0}) = \,\frac{N_F \, e^2}{8 \,\I q_0}\,\left( 1 - \frac{\al}{4} \right)\, .
\label{K1-res}
\eea
Combining Eqs.~(\ref{sigma-cc2}) and (\ref{K1-res}), we arrive, once again, at: $\tilde{\sigma}_1(q_0) = \sigma_0\,(1 + \tilde{\mathcal{C}}'^{(D)} \al + \Ord(\al^2))$, with $\tilde{\mathcal{C}}'^{(D)} = -1/4$.

\begin{widetext}

{\it Optical conductivity from Eq.~(\ref{sigma-dd}) -} We now proceed on computing the 2-loop corrections displayed on Fig.~\ref{fig:2loop-polarization}:
$\Pi_2^{\mu \nu}(q) = 2 \Pi_{2a}^{\mu \nu}(q) + \Pi_{2b}^{\mu \nu}(q)$
where $\Pi_{2a}$ is the so-called self-energy correction and $\Pi_{2b}$ is the so-called vertex correction. The latter are defined as:
\begin{subequations}
\label{def:pi2loop}
\bea
&&\I \Pi_{2a}^{\mu \nu}(q) =
- \int [\D^{d_e}k ]\, \Tr \left[ (-\I e\gamma^\nu) \,S_0(k+q)\,(-\I e\gamma^\mu)\, S_0(k)\left( -\I \Sigma_1(k) \right)\,S_0(k) \right]\, ,
\label{def:pi2a} \\
&&\I \Pi_{2b}^{\mu \nu}(q) = - \int [\D^{d_e}k_1][\D^{d_e}k_2]\,
\Tr \left[ (-\I e\gamma^\nu)\, S_0(k_2+q)\, (-\I e\gamma^0)\, S_0(k_1+q)\, (-\I e\gamma^\mu)\, S_0(k_1)\, (-\I e\gamma^0)\, S_0(k_2)\,
V_{0}(k_1-k_2) \right]\, .
\eea
\end{subequations}
Let's first focus on $\Pi_{2a}^{0 0}$. Performing the trace, going to euclidean space, integrating over frequencies, taking the $\vec{q} \ra 0$ limit 
and substituting the expression of the fermion self-energy Eq.~(\ref{fsigma2}), yields:
\be
\Pi_{2a}^{00}(q_{E0},\vec{q} \ra 0) = - \frac{N_F}{32}\,|\vec{q}\,|^2\,\frac{e_0^4}{v_0^2\,(4\pi)^{D_e/2}}\,\frac{D_e-1}{D_e}\,G(1/2,1/2)\,
\int [\D^{D_e}k ]\,
\frac{|\vec{k}\,|^2 - m_0^2}{[|\vec{k}\,|^2]^{1/2+\veps_\gamma}\, [|\vec{k}\,|^2 + m_0^2]^2}\, .
\label{pi2a-inter}
\ee
The integral is again of the semi-massive one-loop tadpole type and is straightforwardly computed, so:
\be
\Pi_{2a}^{00}(q_{E0},\vec{q} \ra 0) = - \frac{N_F}{32}\,|\vec{q}\,|^2\,\frac{e_0^4\,(m^2)^{D_e/2-3/2-\veps_\gamma}}{v^2\,(4\pi)^{D_e}}\,\frac{(D_e-1)\,(D_e-2-2\veps_\gamma)}{D_e}\,G(1/2,1/2)\,B(1,1/2+\veps_\gamma)\, ,
\ee
\end{widetext}
where we have changed $v_0 \ra v$, with two-loop accuracy. Hence:
\be
2\,\Pi_{2a}^{00}(q_0,\vec{q} \ra 0) = - \frac{N_F\, e^2}{8}\,\frac{\al}{2}\,\frac{|\vec{q}\,|^2}{\I q_0}\, .
\label{pi2a00}
\ee
Combining Eqs.~(\ref{sigma-dd}) and (\ref{pi2a00}), we arrive at: $\sigma_{2a}(q_0) = \sigma_0\,\mathcal{C}_a^{(D)} \al + \Ord(\al^2)$,
and we recover the result of JVH: $\mathcal{C}_{a}^{(D)} =1/2$. Proceeding in a similar way for the vertex correction, the latter can be written in diagrammatic form:
\bea
&&\Pi_{2b}^{0 0}(m,\vec{q} \ra 0\,) = \frac{N_F\,e^4}{8\,v^2}\,\frac{|\vec{q}\,|^2}{D_e}\, 
\times
\biggl\{
(D_e-1)\, I_1(1/2)  \biggr .
\nonum \\
&&\biggl . - m^2 \, I_2(3/2)-m^2 (D_e-2)\, I_0(1/2) \biggr\}\, ,
\label{pi2b00-inter}
\eea
with 2-loop accuracy. Using the results of Eqs.~(\ref{master-dd}), yields:
\be
\Pi_{2b}^{00}(q_{0},\vec{q} \ra 0) = -\frac{N_F \,e^2}{8}\,\al\,\frac{8-3 \pi}{6}\,\frac{|\vec{q}\,|^2}{\I q_0}\, .
\label{pi2b00}
\ee
Combining Eqs.~(\ref{sigma-dd}) and (\ref{pi2b00}), we arrive at: $\sigma_{2b}(q_0) = \sigma_0\,\mathcal{C}_b^{(D)} \al + \Ord(\al^2)$,
and we recover the result: $\mathcal{C}_b^{(D)} =  (8-3\pi)/6$. Adding the two-loop contributions, 
$\sigma_{2}(q_0) =\sigma_{2a}(q_0) + \sigma_{2b}(q_0) = \sigma_0\,\mathcal{C}^{(D)} \al + \Ord(\al^2)$, 
we recover the result of JVH, Eq.~(\ref{c-D}). 
Adding moreover the contribution of the one loop counter-term, $\mathcal{C}'^{(D)}=-1/4$, the total conductivity, with two loop accuracy, reads: 
$\sigma(q_0) = \sigma_{1}(q_0) + \sigma_{2}(q_0) = \sigma_0\,(1+\mathcal{C}^{(DR)} \al + \Ord(\al^2))$, and we finally arrive at the advertised result, Eq.~(\ref{c-DR}).

{\it Optical conductivity from Eq.~(\ref{sigma-cc2}) -} We proceed from Eq.~(\ref{sigma-cc2}) with $K_2(q_0) = 2K_{2a}(q_0) + K_{2b}(q_0)$ 
which should be properly expressed in terms of renormalized parameters. Calculations are done along the same lines as for $\Pi^{00}$.
For the self-energy correction, $K_{2a}$, they yield:
\bea
&&K_{2a}(q_{E0}) = -\frac{N_F}{32 m_0^2 v_0^2}\,\frac{e_0^4}{(4\pi)^{D_e/2}}\,\frac{D_e-1}{D_e}\,G(1/2,1/2)\,
\nonum \\
&&\times \int [\D^{D_e}k]\,\frac{|\vec{k}\,|^2 - m_0^2}{[|\vec{k}\,|^2]^{-1/2+\veps_\gamma}\,[|\vec{k}\,|^2 + m_0^2]^2}\, ,
\label{pi2amumu-inter}
\eea
where the integral is again of the semi-massive tadpole type. Contrary to the case of $\Pi^{00}$, however, it is divergent:
\bea
&&K_{2a}(q_{E0}) = - \frac{N_F}{32 m}\,\frac{e_0^4\,(m^2)^{-2\veps_\gamma}}{v^2\,(4\pi)^{D_e}}\, \frac{(D_e-1)(D_e-2\veps_\gamma)}{D_e}\,
\nonum \\
&&\times \, G(1/2,1/2)\,B(1,-1/2+\veps_\gamma)\, ,
\eea
where, to 2-loop accuracy, $v_0 \ra v$. Performing the $\veps_\gamma$-expansion yields:
\be
2K_{2a}(q_{0}) = \frac{N_F\,e^2}{8\I q_0}\,\frac{\al}{4}\,\left( -\frac{1}{\veps_\gamma} + 2 L_q + 3 - 4 \log 2 + \Ord(\veps_\gamma) \right)\, .
\label{pi2amumu-expr}
\ee
Similarly, after lengthy calculations, 
the vertex correction, $K_{2b}$, reads, in diagrammatic form:
\bea
&&K_{2b}(q_{E0}) = \frac{N_F}{8\,m_0^2}\,\frac{e^4}{v_0^2}\,\frac{1}{D_e} \times
\biggl\{-(D_e-1)\,m^2\,I_1(1/2) \biggr .
\nonum \\
&&\biggl . + I_2(1/2) + (D_e-2)\,I_0(-1/2) \biggr\}\, .
\label{pi2bmumu-inter}
\eea
Using Eqs.~(\ref{master-cc}), yields, with 2-loop accuracy:
\bea
&&K_{2b}(q_{E0}) = \frac{N_F}{8 m}\,\frac{e_0^4\,(m^2)^{-2\veps_\gamma}}{v^2\,(4\pi)^{D_e}}\,
\times \biggl [ \pi \,\left( \pi - \frac{2}{3} \right) \biggr . 
\nonum \\
&&\biggl . - \frac{3-D_e}{D_e}\,G(1/2,1/2)\,B(1,1/2+\veps_\gamma) \biggr]\, .
\eea
Performing the $\veps_\gamma$-expansion, the final result can be put in the form:
\be
K_{2b}(q_0) = -2K_{2a}(q_0)
+\frac{N_F\,e^2}{8\,\I q_{0}}\,\al\,\frac{11 - 3\pi}{6}\,\, .
\label{pi2bmumu-expr}
\ee
Singular terms cancel from the sum of Eqs.~(\ref{pi2amumu-expr}) and (\ref{pi2bmumu-expr}). Combining these equations with Eq.~(\ref{sigma-cc2})
yields: $\tilde{\sigma}_{2}(q_0) = \tilde{\sigma}_{2a}(q_0) + \tilde{\sigma}_{2b}(q_0) = \sigma_0\,\tilde{\mathcal{C}}^{(D)} \al + \Ord(\al^2)$,
 and we recover the result of JVH: 
$\tilde{\mathcal{C}}^{(D)} = (11-3\pi)/6$. Adding the contribution of the one loop counterterm, $\tilde{\mathcal{C}'}^{(D)}=-1/4$, 
the total conductivity, with two loop accuracy, reads:
$\tilde{\sigma}(q_0) = \tilde{\sigma}_{1}(q_0) + \tilde{\sigma}_{2}(q_0) = \sigma_0\,(1+\tilde{\mathcal{C}}^{(DR)} \al + \Ord(\al^2))$,
yielding, once again, the advertised result of Eq.~(\ref{c-DR}).

{\it Conclusion -} The two-loop interaction correction to the minimal conductivity of graphene has been computed in two different ways 
yielding a single result: $\mathcal{C}^{(2)} = (19-6\pi)/12 \approx 0.013$, compatible with experimental uncertainties.~\cite{C-experiments} 
Upon deriving this result, a crucial role was played by Fermi velocity renormalization.~\cite{GonzalezGV93,JuanGV10} 
The latter had to be properly taken into account within dimensional regularization.  
Moreover, we have used algebraic techniques, originally developed for relativistic quantum field theories such as QED and QCD, in order to compute dimensionally regularized Feynman integrals.
By combining such techniques with proper renormalization, our approach yields a general prescription
to systematically compute interaction corrections in pseudo-relativistic systems such as graphene. 
The present results were derived in the non-relativistic case ($v/c \ra 0$).
A similar prescription has been used to study the ultrarelativistic case~\cite{Teber12,KotikovT13} ($v/c \ra 1$)
which corresponds to the infrared fixed point of graphene.~\cite{GonzalezGV93}
Interestingly, at the fixed point: $\mathcal{C}^{*} = (92-9\pi^2)/(18\pi) \approx 0.056$, which is of the same order of magnitude as $\mathcal{C}^{(2)}$.

\acknowledgments
The work of A.V.K.\ was supported in part by the Russian Foundation for Basic
Research grant No. 13-02-01005-a and by the Universit\'e Pierre et Marie Curie (UPMC).
Discussions with Marc Bellon are warmly acknowledged.

\end{document}